\documentclass[twocolumn,times]{aastex6}
\usepackage{jab}
\hypersetup{citecolor=black,linkcolor=black}

\slugcomment{}
\shorttitle{Multi-Species Instability Thresholds}
\shortauthors{Chen et al.}

\begin{document}

\title{Multi-Species Measurements of the Firehose and Mirror Instability\\Thresholds in the Solar Wind}
\author{C. H. K. Chen\altaffilmark{1,2}, L. Matteini\altaffilmark{1}, A. A. Schekochihin\altaffilmark{3,4}, M. L. Stevens\altaffilmark{5}, C. S. Salem\altaffilmark{2}, B. A. Maruca\altaffilmark{2}, M. W. Kunz\altaffilmark{6}, \\and S. D. Bale\altaffilmark{2,7}}
\affil{$^1$Department of Physics, Imperial College London, London SW7 2AZ, UK; christopher.chen@imperial.ac.uk\\
$^2$Space Sciences Laboratory, University of California, Berkeley, California 94720, USA\\
$^3$Rudolf Peierls Centre for Theoretical Physics, University of Oxford, Oxford OX1 3NP, UK\\
$^4$Merton College, Oxford OX1 4JD, UK\\
$^5$Harvard-Smithsonian Center for Astrophysics, Cambridge, Massachusetts 02138, USA\\
$^6$Department of Astrophysical Sciences, Princeton University, Princeton, New Jersey 08544, USA\\
$^7$Department of Physics, University of California, Berkeley, California 94720, USA}
\begin{abstract}
The firehose and mirror instabilities are thought to arise in a variety of space and astrophysical plasmas, constraining the pressure anisotropies and drifts between particle species. The plasma stability depends on all species simultaneously, meaning that a combined analysis is required. Here, we present the first such analysis in the solar wind, using the long-wavelength stability parameters to combine the anisotropies and drifts of all major species (core and beam protons, alphas, and electrons). At the threshold, the firehose parameter was found to be dominated by protons (67\%), but also to have significant contributions from electrons (18\%) and alphas (15\%). Drifts were also found to be important, contributing 57\% in the presence of a proton beam. A similar situation was found for the mirror, with contributions of 61\%, 28\%, and 11\% for protons, electrons, and alphas, respectively. The parallel electric field contribution, however, was found to be small at 9\%. Overall, the long-wavelength thresholds constrain the data well ($<$1\% unstable), and the implications of this are discussed.
\end{abstract}

\section{Introduction}

Many space and astrophysical plasmas are weakly collisional so that their particle distributions can be sufficiently non-Maxwellian to provide a source of free energy for velocity space instabilities. These are thought to arise in a variety of environments, e.g., planetary magnetosheaths \citep{soucek08}, cometary magnetospheres \citep{russell87}, accretion disks \citep{sharma06}, and galaxy clusters \citep{schekochihin08b}, and to affect plasma processes such as thermal conduction \citep{schekochihin08b}, angular momentum transport \citep{sharma06}, and reconnection \citep{matteini13b}. The solar wind, which has significant temperature anisotropies, beams and inter-species drifts, allows the behaviour of these instabilities to be studied in detail, through the use of large in situ data sets.

The three main particle species that make up the solar wind---H$^+$ ions (protons), He$^{2+}$ ions (alpha particles), and electrons---have long been known to display significant temperature anisotropies with respect to the magnetic field direction \citep[e.g.,][]{hundhausen67}, with the ratio of perpendicular to parallel temperature $T_\perp/T_\|$ differing from unity by up to an order of magnitude in both directions. These anisotropies arise from processes such as the solar wind expansion \citep{parker58b}, resonant wave-particle interactions \citep{cranmer14}, pickup ions \citep{isenberg95}, turbulent shear \citep{schekochihin08b}, and other local expansions and compressions. If the anisotropy becomes large enough, the firehose and mirror instabilities \citep{rosenbluth56,parker58b,chandrasekhar58,vedenov58,hasegawa69} can be triggered. At large scales, the firehose instability is fluid in nature and arises when the total parallel pressure is large enough to cause Alfv\'en waves to grow in amplitude. It is present in anisotropic fluid models \citep[e.g.,][]{kunz15}, and is sometimes known as the non-resonant firehose, since it does not involve wave-particle resonances but saturates adiabatically \citep{davidson68} or by particle scattering \citep{kunz14}. The long-wavelength mirror instability, however, is inherently kinetic \citep{southwood93}, involving Landau resonant particles, and saturates via particle trapping \citep{kivelson96}. It arises when compressive fluctuations become unstable \citep{southwood93,klein15b} due to the anisotropy ($T_\perp > T_\|$) of both ions and electrons. A variety of short-wavelength temperature-anisotropy instabilities can also arise, which include the parallel firehose, oblique firehose (the short wavelength extension of the fluid firehose \citep{klein15b}), ion cyclotron, mirror, oblique electron firehose, and electron whistler instabilities \citep[see][for a review]{gary15}.

As well as temperature anisotropies, drifts (differing bulk velocities) between species, and between different populations of the same species, are present in the solar wind \citep{feldman73,marsch82a,marsch82b} and can also drive instabilities. In particular, the protons often consist of ``core'' (higher density) and ``beam'' (lower density) populations, which can differ significantly in their bulk velocities. Drifts may arise from stream mixing \citep{feldman73}, wave-particle interactions \citep{matteini10}, reconnection \citep{gosling05}, coronal heating mechanisms \citep{isenberg83}, or other processes. Similarly to temperature anisotropies, they can induce both large and small scale instabilities. Since each species should have the same ``E$\times$B'' perpendicular velocity, drifts occur parallel to the magnetic field, effectively increasing the parallel pressure, and can excite the fluid firehose instability, even if each population is isotropic \citep{parker61,kunz15}. Resonant Alfv\'en and magnetosonic ion drift instabilities can also arise \citep{daughton98}, as well as electron drift (or heat flux) instabilities \citep{gary75}.

The non-linear evolution of these instabilities is expected to bring the plasma to marginal stability. Solar wind measurements show the anisotropy of protons \citep{gary01,kasper02,hellinger06a,bale09}, alphas \citep{maruca12}, and electrons \citep{stverak08}, as well as alpha-proton \citep{marsch87} and proton core-beam \citep{marsch82a,marsch87} drifts, to be mostly constrained to stable values, consistent with this hypothesis. A difficulty faced by previous analyses, however, is the large number of parameters involved: plasma stability depends on all species simultaneously. While this has been partially addressed by some observational studies \citep{dum80,maruca12,verscharen13b,bourouaine13}, a complete analysis based on all major parameters has not previously been performed. In this Letter, we present the first such analysis, which includes both temperature anisotropies and drifts of all major species (core and beam protons, alphas, and electrons), to investigate the firehose and mirror instabilities.

\section{Data Set}

For the analysis, data from the \emph{Wind} spacecraft at 1 AU were used. The ion data, derived from the SWE instrument \citep{ogilvie95}, consist of a three-population fit to each Faraday cup spectrum: a bi-Maxwellian core proton population, a Maxwellian proton beam population, and a bi-Maxwellian alpha population. The fit allows the populations to drift freely, except for the core-beam drift, which is constrained to be parallel to the magnetic field, as measured by MFI \citep{lepping95}. A proton beam was determined to be present if the resulting $\chi^2$ per degree of freedom with the extra three parameters (beam density, speed, and temperature) was more favourable, which occurrs 30\% of the time. The electron data, derived from the 3DP electrostatic analysers \citep{lin95}, consist of moments of the measured distributed function, after correction for spacecraft charging, photoelectrons, and other effects \citep{pulupa14}. As part of the correction process, the electron density was constrained by the value determined from the thermal noise measurements.

The integration time is $\sim$90\,s for each ion distribution and $\sim$3\,s for each electron distribution, and the time between measurements is $\sim$90\,s for ions and $\sim$100\,s for electrons (not all electron distributions are telemetered), meaning that the ion and electron data are not aligned in time. They were merged by selecting pairs of ion and electron measurements that differ in time by up to 20\,s (the results are not sensitive to this value) so that the corresponding electron distribution is well within the the ion measurement. The resulting data set consists of 150,981 points covering four years (1995-1998) when the spacecraft was in the solar wind. Since the ion fits are sometimes unreliable, cuts were made to the data based on visual inspection of the parameter distributions; in most cases there is a clear distinction between the physical distribution and erroneous fits. After these cuts, 108,099 points remain, which were used for the work presented here.

\section{Results}

\subsection{Pressure Anisotropy}

\begin{figure}
\includegraphics[width=\columnwidth,trim=0 0 0 0,clip]{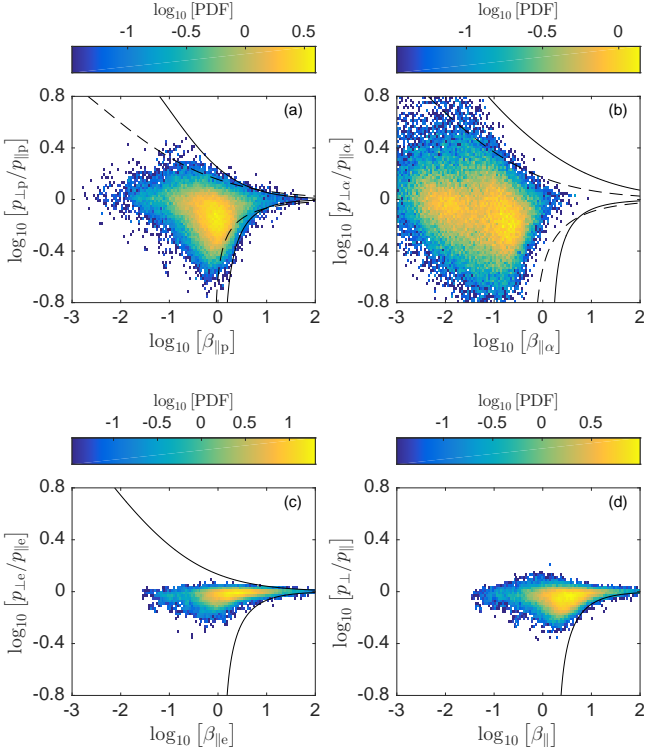}
\caption{2D PDFs of the pressure anisotropy and parallel beta for (a) protons, (b) alphas, (c) electrons, and (d) combined species, for times when a proton beam was not present. The solid and dashed lines correspond to various instability thresholds (see text for details).}
\label{fig:anisotropybeta}
\end{figure}

Figure \ref{fig:anisotropybeta}(a-c) shows the 2D probability density functions (PDFs) of pressure anisotropy $p_{\perp s}/p_{\|s}$ and parallel beta $\beta_{\|s}$ for each species $s$ during times when a proton beam was not present. Consistent with previous studies \citep[e.g.,][]{hellinger06a,stverak08,maruca12}, they are mostly constrained to the stable sides of instability boundaries determined from numerical solution of the hot plasma dispersion relation. The thresholds marked in Figure \ref{fig:anisotropybeta} are (a) the proton mirror (upper solid), proton cyclotron (upper dashed), proton oblique firehose (lower solid), and proton parallel firehose (lower dashed) from \citet{hellinger06a}; (b) the alpha mirror (upper solid), alpha cyclotron (upper dashed), alpha oblique firehose (lower solid), and alpha parallel firehose (lower dashed) from \citet{maruca12}; and (c) the electron whistler (upper) from \citet{gary96} and electron firehose (lower) from \citet{hellinger14a}. The fact that the distribution in Figure \ref{fig:anisotropybeta}(a) does not reach as close to the mirror instability threshold as previous studies \citep{hellinger06a,bale09} is due to the smaller data set used here.

The thresholds in Figure \ref{fig:anisotropybeta}(a-c), however, are not complete, since in each one all other species are assumed isotropic. To investigate the anisotropy of all species together, the PDF of the total pressure anisotropy $p_\perp/p_\|$ and $\beta_\|$, where $p_{\perp,\|}=\sum_sp_{\perp,\| s}$, and $\beta_{\perp,\|}=\sum_s\beta_{\perp,\| s}$, is shown in Figure \ref{fig:anisotropybeta}(d). Also marked is the long-wavelength firehose threshold $\beta_\|-\beta_\perp =2$. It can be seen that the distribution is well constrained to the stable side and its contours follow the shape of the threshold. This suggests that the non-resonant firehose instability may be playing a role in constraining the multi-species solar wind evolution.

\subsection{Firehose Instability}

\begin{figure}
\includegraphics[width=\columnwidth,trim=0 0 0 0,clip]{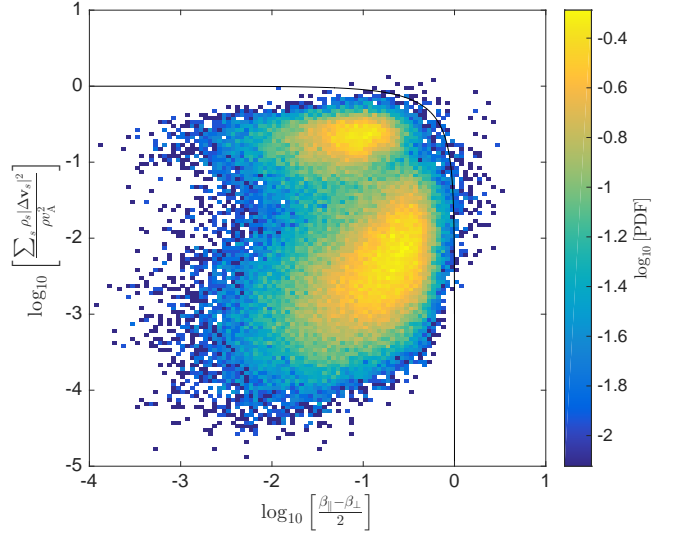}
\caption{2D PDF of the anisotropy term and drift term of the firehose instability threshold (Equation (\ref{eq:fhwithdrifts})). The black line marks the  threshold, with the majority of the distribution on the stable side. Two populations can be seen, corresponding to whether a proton beam is present (upper population) or not (lower population).}
\label{fig:anisotropydrift}
\end{figure}

The condition for the long-wavelength firehose instability in a plasma with both anisotropies and drifts is \citep[e.g.,][]{kunz15}
\begin{equation}
\Lambda_\mathrm{f}\equiv\frac{\beta_\|-\beta_\perp}{2}+\frac{\sum_s\rho_s|\Delta\mathbf{v}_s|^2}{\rho v_\mathrm{A}^2}>1,
\label{eq:fhwithdrifts}
\end{equation}
where $\rho_s$ is the mass density of species $s$, $\rho$ is the total mass density, $v_\mathrm{A}$ is the Alfv\'en speed, and $\Delta\mathbf{v}_s$ is the difference between the bulk velocity of species $s$ and the centre of mass velocity $\mathbf{v}$. Figure \ref{fig:anisotropydrift} shows the distribution of the two terms of $\Lambda_\mathrm{f}$; the total pressure anisotropy and the sum of the drifts (times when the anisotropy is negative are excluded). The distribution appears to be constrained to the stable region due to both anisotropy and drifts as well as a combination of the two. There also appears to be two populations, which correspond to whether a proton beam is present or not: the population near the drift threshold (upper population) corresponds to the presence of a beam and the population near the anisotropy threshold (lower population) to the absence of a beam.

\begin{figure}
\includegraphics[width=\columnwidth,trim=0 0 0 0,clip]{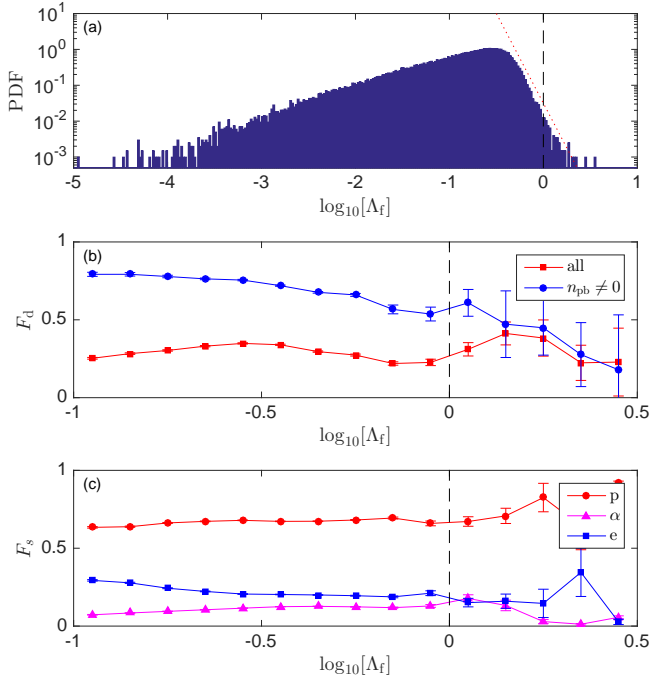}
\caption{(a) PDF of the firehose parameter $\Lambda_\mathrm{f}$ (Equation (\ref{eq:fhwithdrifts})). (b) Mean fractional contribution $F_\mathrm{d}$ of drifts (all species) to $\Lambda_\mathrm{f}$ for the whole data set (red squares) and when proton beams are present (blue circles). (c) Mean fractional contribution $F_s$ of species $s$ (both anisotropy and drifts) to $\Lambda_\mathrm{f}$. The black dashed lines mark the instability threshold and the red dotted line in (a) is a slope of gradient $-5$.}
\label{fig:fhfrac}
\end{figure}

The distribution of $\Lambda_\mathrm{f}$ (for $\Lambda_\mathrm{f}>0$) is shown in Figure \ref{fig:fhfrac}(a) (data with $\beta>10$ were excluded due to the large error on $\Lambda_\mathrm{f}$ introduced by a small error on the anisotropy). While the shape of the distribution depends on the particular form of the instability parameter, the fraction on either side of the threshold can be meaningfully compared. Only 0.1\% is in the unstable region ($\Lambda_\mathrm{f}>1$), consistent with the firehose instability boundary being a constraint on the combined anisotropies and drifts of all species in the solar wind.

Since both the species contributions and the drift and anisotropy terms are additive in Equation (\ref{eq:fhwithdrifts}), the fractional contributions of each can also be determined. Figure \ref{fig:fhfrac}(b) shows the binned and averaged fractional contributions of the drift term to $\Lambda_\mathrm{f}$, as a function of $\Lambda_\mathrm{f}$, for the whole data set and for times when a proton beam was present. The error bars represent the standard error of the mean. At the threshold, $\Lambda_\mathrm{f}=1$, the drifts contribute 27\% overall, but this rises to 57\% in the presence of a proton beam. Figure \ref{fig:fhfrac}(c) shows the contribution of each species to $\Lambda_\mathrm{f}$, where at the threshold protons contribute 67\%, electrons 18\%, and alphas 15\%. Therefore, while the protons are dominant, non-proton species contribute around one third to the instability of the plasma.

\subsection{Mirror Instability}

\begin{figure}
\includegraphics[width=\columnwidth,trim=0 0 0 0,clip]{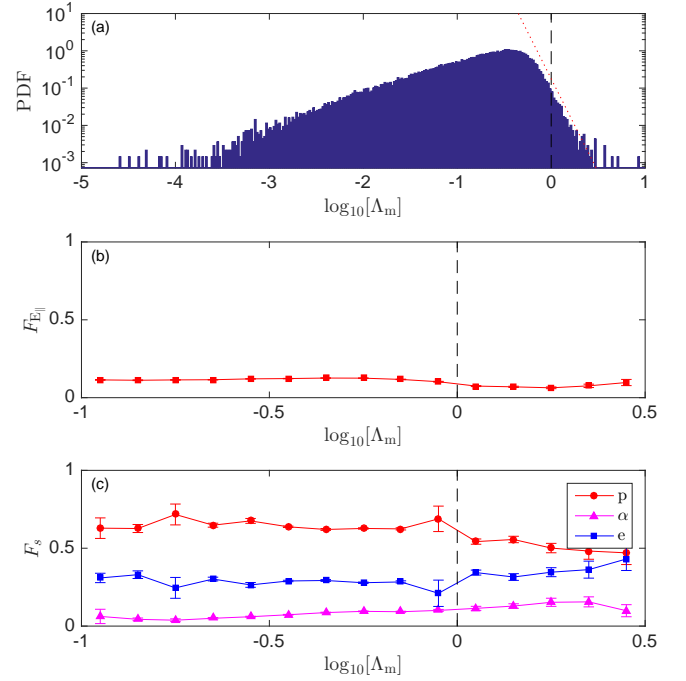}
\caption{(a) PDF of the mirror parameter $\Lambda_\mathrm{m}$ (Equation (\ref{eq:mirror})). (b) Mean fractional contribution $F_\mathrm{E_\|}$ of the parallel electric field to $\Lambda_\mathrm{m}$. (c) Mean fractional contribution $F_s$ of species $s$ to $\Lambda_\mathrm{m}$. The black dashed lines mark the instability threshold and the red dotted line in (a) is a slope of gradient $-5$.}
\label{fig:mfrac}
\end{figure}

The condition for the long-wavelength mirror instability is given by \citep[e.g.,][]{hellinger07}
\begin{equation}
\label{eq:mirror}
\Lambda_\mathrm{m}\equiv\sum_s\beta_{\perp s}\left(\frac{\beta_{\perp s}}{\beta_{\| s}}-1\right)-\frac{\left(\sum_sq_sn_s\frac{\beta_{\perp s}}{\beta_{\| s}}\right)^2}{2\sum_s\frac{(q_sn_s)^2}{\beta_{\| s}}}>1,
\end{equation}
where $q_s$ is the charge of species $s$. The distribution of $\Lambda_\mathrm{m}$ (for $\Lambda_\mathrm{m}>0$) is shown in Figure \ref{fig:mfrac}(a), in which times when a proton beam was present were excluded, since Equation (\ref{eq:mirror}) is for a plasma without drifts (as for the firehose, data with $\beta>10$ were also excluded). Again, only a small fraction of the data (0.7\%) are unstable ($\Lambda_\mathrm{m}>1$), consistent with the mirror instability constraint on the total anisotropy of all species in the plasma. The second term of $\Lambda_\mathrm{m}$ in Equation (\ref{eq:mirror}) originates from the parallel electric field $E_\|$, and its fractional contribution to $\Lambda_\mathrm{m}$ is shown in Figure \ref{fig:mfrac}(b). It can be seen that this is consistently small, with a value of 9\% at $\Lambda_\mathrm{m}=1$. Neglecting this term in Equation (\ref{eq:mirror}), the fractional contribution of each species to $\Lambda_\mathrm{m}$ can be determined, and this is shown in Figure \ref{fig:mfrac}(c). Similarly to the firehose, protons are dominant (61\%) at $\Lambda_\mathrm{m}=1$, but electrons (28\%) and alphas (11\%) together contribute around one third to the the instability of the plasma.

\section{Discussion}

\begin{figure}
\includegraphics[width=\columnwidth,trim=0 0 0 0,clip]{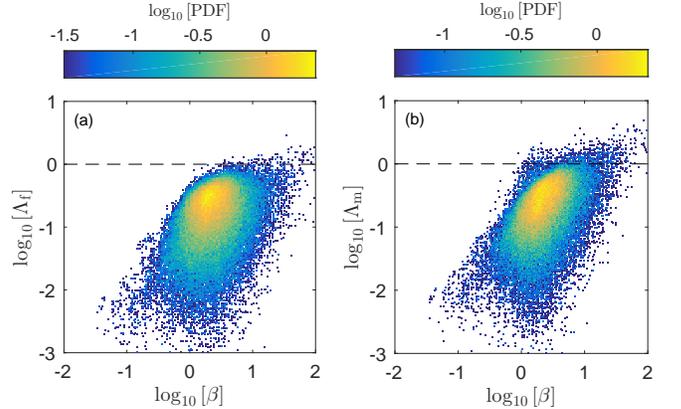}
\caption{(a) 2D PDF of the firehose instability parameter $\Lambda_\mathrm{f}$ and the total plasma beta $\beta$. (b) Same for the mirror instability parameter $\Lambda_\mathrm{m}$. The black dashed lines mark the instability thresholds.}
\label{fig:lambdabeta}
\end{figure}

Through the use of analytic thresholds for the long-wavelength firehose and mirror instabilities, we have been able to examine the combined constraints they impose on the major solar wind species. Figures \ref{fig:fhfrac} and \ref{fig:mfrac} show that for the majority of time, the solar wind is stable, although there is not a sharp cut off in the instability parameters at the thresholds. There is an exponential drop in the PDF of $\log [\Lambda_\mathrm{f,m}]$, which can be interpreted as being due to the balance between the processes that generate anisotropy, such as the turbulent fluctuations, and those, i.e., the instabilities, that reduce it. These instabilities are expected to be important in high $\beta$ plasma, and this is confirmed in Figure \ref{fig:lambdabeta}, which shows the 2D distribution of $\Lambda_\mathrm{f,m}$ and $\beta$, indicating that for the solar wind at 1 AU, the thresholds are reached for $\beta\gtrsim 1$.

The protons were found to dominate both instability parameters at their thresholds, although the other species were found to contribute around one third to each, making their inclusion important when using such thresholds to determine which instabilities are active. While here we have considered the long-wavelength instabilities, similar considerations apply to the variety of short-wavelength temperature anisotropy and drift instabilities. An interesting question is why the data reaches the firehose and mirror thresholds when those of other kinetic instabilities, such as the parallel (whistler) firehose and ion cyclotron, have been suggested to be lower \citep[e.g.,][]{hellinger06a}. To address this with solar wind data, it is important to consider the multi-species nature of the plasma, in addition to the details of the particle distributions \citep{hellinger08a,isenberg13}, which will require a numerical treatment.

Finally, the results of this Letter may be useful for large-scale transport models of weakly collisional astrophysical plasmas. For example, models of both the solar wind \citep{chandran11} and radiatively-inefficient accretion flows \citep{sharma06} have employed instability thresholds to constrain the evolution of the pressure anisotropy. The observations in this Letter show that the long-wavelength firehose and mirror thresholds provide good constraints when multiple anisotropic drifting species are present.

\acknowledgments
C.H.K.C. is supported by an Imperial College Junior Research Fellowship. L.M. is supported by STFC Grant ST/K001051/1. We acknowledge support provided to ISSI/ISSI-BJ Team 304 and the Marie Curie Project FP7 PIRSES-2010-269297 -- ``Turboplasmas''. We thank P. Hellinger, K. G. Klein, and D. Stansby for useful discussions.

\bibliography{bibliography}

\end{document}